\documentclass[trackchanges, twocolumn]{aastex701}
\def\hi{H{\sc i}}
\def\hhh{H$_3^+$}
\def\arcsec{\hbox{$^{\prime\prime}$}}
\def\arcmin{\hbox{$^{\prime}$}}
\def\deg{$^\circ$}
\def\cm2{cm$^{-2}$}
\def\cc{cm$^{-3}$}
\def\kms{km s$^{-1}$}
\def\s{s$^{-1}$}
\def\nh3{NH$_3$}
\def\n2h{N$_2$H$^+$}
\def\co{$^{12}$CO}
\def\13co{$^{13}$CO}
\def\c18o{C$^{18}$O}
\def\hc3n{HC$_3$N}
\def\h2{H$_2$}
\def\nh{n(H$_2$)}
\def\c2{[C\,{\sc ii}]}

\begin{document}

\title{Star Formation Drives Production of Low Energy Cosmic Rays}

\author[orcid=0000-0002-2169-0472]{Ningyu Tang}
\affiliation{Department of Physics, Anhui Normal University, Wuhu 241002, China}
\email{nytang@ahnu.edu.cn}

\author{Jiahao Liu} 
\affiliation{Department of Astronomy, University of Science and Technology of China, Hefei 230026, China}
\email{ljhstpc11@mail.ustc.edu.cn}

\author[ orcid=0000-0003-3010-7661]{Di Li}
\altaffiliation{Corresponding author: Di Li}
\affiliation{New Cornerstone Science Laboratory, Department of Astronomy, Tsinghua University, Beijing 100084, China}
\affiliation{National Astronomical Observatories, Chinese Academy of Sciences, Beijing 100101, China}
\email[show]{dili@tsinghua.edu.cn}

\author{Ruizhi Yang }
\affiliation{Department of Astronomy, University of Science and Technology of China, Hefei 230026, China}
\email{yangrz@ustc.edu.cn}

\author[0000-0003-2733-4580]{Thomas G. Bisbas}
\affiliation{ Research Center for Computational Earth and Space Science, Zhejiang Lab, Hangzhou 311100, People’s Republic of China }
\email{tbisbas@zhejianglab.com }

\author{ Bing Liu}
\affiliation{ Division of Dark Matter and Space Astronomy, Purple Mountain Observatory, Chinese Academy of Sciences, Nanjing 210023, People’s Republic of China }
\email{liubing@pmo.ac.cn }

\author{Marko Kr\v{c}o }
\affiliation{ National Astronomical Observatories, Chinese Academy of Sciences, Beijing 100101, China }
\email{marko@nao.cas.cn }

\author{ Paul Goldsmith}
\affiliation{ Jet Propulsion Laboratory, California Institute of Technology, 4800 Oak Grove Drive, Pasadena, CA 91109, USA }
\email{ Paul.F.Goldsmith@jpl.nasa.gov}

\author{ Paola Caselli }
\affiliation{ Max-Planck-Institut für Extraterrestrische Physik, D-85748 Garching, Germany }
\email{ caselli@mpe.mpg.de}

\author{Sihan Jiao }
\affiliation{Max Planck Institute for Astronomy, Konigstuhl 17, D-69117 Heidelberg, Germany  }
\email{ sihanjiao@nao.cas.cn}

\author[0000-0002-3866-414X]{Yan Gong }
\affiliation{ Purple Mountain Observatory and Key Laboratory of Radio Astronomy, Chinese Academy of Sciences,Nanjing 210023, People’s Republic of China }
\email{  ygong@pmo.ac.cn}

\author{ Gan Luo}
\affiliation{ Institut de Radioastronomie Millimetrique, 300 rue de la Piscine, 38400, Saint-Martin d’Hères, France }
\email{luo@iram.fr }

\author[orcid=0000-0002-7020-4290]{Xinwen Shu }
\affiliation{Department of Physics, Anhui Normal University, Wuhu 241002, China}
\email{ xwshu@ahnu.edu.cn}

\author[orcid=0009-0000-8497-8476]{Liangchong Zhu}
\affiliation{Department of Physics, Anhui Normal University, Wuhu 241002, China}
\email{lczhu@ahnu.edu.cn}

\author{Xiaohui Sun }
\affiliation{School of Physics and Astronomy, Yunnan University, Kunming 650500, China  }
\email{xhsun@ynu.edu.cn }

\author{Chen Wang }
\affiliation{ Institute of Astronomy and Information, Dali University, Dali 671003, China }
\email{wangchen@pmo.ac.cn }

\author[0000-0001-8516-2532]{ Tao-Chung Ching}
\affiliation{ National Radio Astronomy Observatory, 1003 Lopezville Road, Socorro, NM 87801, USA }
\email{ tching@nrao.edu}

\author{Donghui Quan  }
\affiliation{ Department of Physics, Xi’an Jiaotong-Liverpool University, 111 Ren’ai Road, Suzhou 215123, China }
\email{donghui.quan@xjtlu.edu.cn }

\author{Junzhi Wang  }
\affiliation{ Guangxi Key Laboratory for Relativistic Astrophysics, Department of Physics,Guangxi University, Nanning 530004, China  }
\email{  junzhiwang@gxu.edu.cn}

\author{ Xuejian Jiang}
\affiliation{ Research Center for Computational Earth and Space Science, Zhejiang Lab, Hangzhou 311100, People’s Republic of China }
\email{  jiangxuejian@zhejianglab.edu.cn}

\author{Pei Zuo }
\affiliation{ National Astronomical Observatories, Chinese Academy of Sciences, Beijing 100101, China }
\email{peizuo@nao.cas.cn }

\begin{abstract}

For over a century, the origin of low-energy cosmic rays (LECRs), the dominant heaters and ionizers of dense interstellar gas, remains elusive owing to solar modulation and uncertain transport processes. 
In this study, we introduce a new astrophysical approach based on \hi\ Narrow Self-Absorption (HINSA) to obtain spatially resolved measurements of LECR ionization rates using high-fidelity \hi\ observations toward the Orion region from the FAST telescope. The LECR ionization rate is found to scale with local star formation rate (SFR) as $log_{10}\zeta = (1.4\pm 0.70)log_{10}\mathrm{SFR} + (-10.5\pm 2.9)$. Moreover, it increases with visual extinction, and is found to exceed, toward active star-forming regions, the value predicted for diffuse regions based on \textit{Voyager} measurements and an external propagation model. These findings demonstrate that LECRs are generated in situ by star-forming activities rather than penetrating from the broader Galactic cosmic-ray population. This is further supported by \textit{Fermi}-LAT gamma-ray observations toward the Orion region. Together, these results resolve a key uncertainty in cosmic-ray origin and establish a new avenue for quantifying the energetic feedback that regulates the interstellar medium.

\end{abstract}



\section{Introduction} 
\label{sec:intro}

Cosmic rays (CRs) are a key ingredient of the interstellar medium (ISM) \citep{2001RvMP...73.1031F}. The power-law energy spectrum of CRs spans many orders of magnitude, dictating only low-energy CRs' (LECRs) having appreciable flux to affect cosmic baryon evolution \citep{ 1968ApJ...152..971S, 1969ApJ...155L.149F}. LECRs initiate key processes in the chemical evolution of the ISM. While Galactic supernovae are well-established sources of high-energy CRs ($E > 1$ GeV), the origin of LECRs remains unresolved \citep{2022A&ARv..30....4G,2026ESC....10..276G}. 

Measurements of LECRs near Earth are heavily influenced by the magnetized solar wind and complex transport effects \citep{2013LRSP...10....3P}. The ionization rate of molecular hydrogen (\h2) by LECRs (LECRIR) serves as a proxy for the flux of LECRs interacting with the ISM \citep{2009A&A...501..619P}. With an ionization potential of 15.4 eV, molecular hydrogen (\h2)  can be ionized by LECRs through the following process:

\begin{equation}
\rm H_2 + CR \rightarrow H_2^+ + e^- + CR,  
\end{equation}
 which is followed by the fast ion-neutral reaction, 
 \begin{equation}
\rm H_2^+ + H_2 \rightarrow H_3^+ + H.  
\label{eq:h3_form}
\end{equation}

Protonated hydrogen, $\mathrm{H}_3^+$  is  a pivotal species in ISM chemistry. It plays a crucial role in various chemical reactions and serves as an important tracer of the physical and chemical conditions in interstellar clouds.  The balance between formation and destruction rates determine final  abundance of $\mathrm{H}_3^+$ under equilibrium.  In diffuse and translucent clouds with relatively large electron fraction $x_e$, e.g.,  $1.6\times 10^{-4}$ at small visual extinctions \citep{2004ApJ...605..272S},  \hhh\ is mainly destroyed by  dissociative recombination:  
 \begin{equation}
\rm H_3^+ + e^- \rightarrow H_2 + H\  or\ H + H + H.  
\label{eq:h3_e_diss}
\end{equation}
In dense clouds where the electron fraction is  small ( $x_e \approx 10^{-7}$) due to effective shielding from UV interstellar radiation, the dissociation-recombination progress becomes unimportant. Instead, the primary destruction pathway of $\mathrm{H}_3^+$ is the reaction with the abundant CO molecule:  
 \begin{equation}
\rm H_3^+ + CO \rightarrow HCO^+ + H_2.  
\label{eq:h3_co}
\end{equation}

Based on the aforementioned reactions, the abundance of \hhh\ is linked to the cosmic ray ionization rate of \h2\ (denoted as  $\zeta$, in both diffuse and dense clouds. Various surveys have been conducted to study \hhh\ in diffuse clouds \citep{1998Sci...279.1910M, 2002ApJ...567..391M, 2003Natur.422..500M, 1999ApJ...510..251G, 2007ApJ...671.1736I, 2010ApJ...724.1357I, 2010ApJ...715..757G, 2011ApJ...729...15C, 2012ApJ...745...91I, 2014ApJ...787...44A} and dense clouds \citep{1996Natur.384..334G, 1999ApJ...522..338M, 2004ApJ...606..911B, 2010ApJ...715..757G, 2015ApJ...806...57G, 2019A&A...632A..29G} toward the spiral arms of the Milly Way.  These surveys suggest a $\zeta$ value on the order of $10^{-16}$ s$^{-1}$ in diffuse clouds, such as an average of  $\sim 3.5 \times 10^{-16}$ $\mathrm{s}^{-1}$ in  \citet{2012ApJ...745...91I},  and a $\zeta$ value on the order of $10^{-17}$ s$^{-1}$ in dense clouds \citep{2000A&A...358L..79V}. Notably, recent studies have shown that downward revisions of the gas density reduce the inferred $\zeta$ values in diffuse clouds by an average factor of $\sim 9$ \citep{2024ApJ...973..143N,2024ApJ...973..142O,2026ApJ...997..123I}. Meanwhile, \citet{2000A&A...358L..79V} pointed out that contamination from \hhh\ in the diffuse envelopes surrounding dense clouds can lead to systematically overestimated $\zeta$ values in dense regions. Starting from the formation of \hhh,  deuterated molecular ions such as $\rm H_2D^+$ and DCO$^+$ can be generated through deuterium fractionation reactions,  $\rm H_3^+ + HD \leftrightarrow H_2D^+ + H_2$ and  $\rm H_2D^+ + CO \rightarrow DCO^+ + H_2 \ or \  HCO^+ + HD$.  When compared to \hhh\ observations,  observations of  $\rm H_2D^+$, HCO$^+$, and DCO$^+$ result in much lower values of $\zeta$ \citep{1998ApJ...499..234C, 2000A&A...358L..79V, 2007ApJ...664..956M, 2008ApJ...684.1221H, 2014A&A...563A.127M, 2016A&A...593A..94F, 2020A&A...644A..34S}.

Oxygen chemistry is initiated by the cosmic-ray ionization of atomic hydrogen, followed by a slightly endothermic charge-exchange reaction with atomic oxygen that produces O$^+$. The resulting O$^+$ then reacts exothermically with H$_2$ to form oxygen-bearing molecular ions such as OH$^+$, H$_2$O$^+$, and H$_3$O$^+$, which can serve as tracers of the cosmic-ray ionization rate $\zeta$ \citep{2012ApJ...754..105H}.
Surveys conducted with Herschel telescope and the very large telescope (VLT) have found   $\zeta$ values  \citep[e.g.,][]{2010A&A...518L.110G, 2010A&A...521L..10N, 2015ApJ...800...40I, 2019A&A...622A..31B}  consistent with those obtained from \hhh\ observations  \citep[e.g.,][]{2012ApJ...745...91I} in diffuse regions.

The current sample of $\zeta$ determinations remains small, with fewer than 80 measurements, primarily due to the high demands of UV/terahertz absorption spectroscopy toward bright background sources and the low abundances of these tracers. Moreover, while H$_3^+$ provides reliable LECRIR measurements, the interpretation of other tracers, such as OH$^+$, depends on a detailed understanding of their complex chemical evolution. A more reliable model-independent LECRIR tracer with high abundance is needed to enable more extensive measurements of $\zeta$ across diverse environments.

While recent observations with the James Webb Space Telescope (JWST) have successfully detected cosmic-ray-induced \h2\ excitation in Barnard 68 \citep{2026NatAs..10..540B, 2026ApJ...998...71N}, direct observations of excited \h2\ remain considerably more challenging because they require extremely high-sensitivity infrared measurements. 

The chemical reaction in Equation~\ref{eq:h3_form} highlights the dominant production pathway of atomic hydrogen (\hi) inside dense molecular clouds, in which \hi\ is produced as a by-product of H$_3^+$ formation initiated by cosmic-ray ionization. Within dense, cold molecular regions, this \hi\ manifests as an absorption feature with a narrow velocity dispersion, typically with FWHM linewidths smaller than 2 km s$^{-1}$. This phenomenon, known as \hi\ Narrow Self-Absorption (HINSA; \citealp{2003ApJ...585..823L, 2005ApJ...622..938G, 2008ApJ...689..276K, 2010ApJ...724.1402K, 2018ApJ...867...13Z, 2020RAA....20...77T}), is commonly spatially correlated with CO and OH emission. Crucially, HINSA traces the cold \hi\ fraction produced via \hhh\  dissociative recombination, which is driven by the ionization of \h2\ by low-energy cosmic rays (LECRs). Given its high relative abundance ($\sim 10^{-3}$) and its direct physical dependency on the ionization rate, HINSA serves as a powerful, large-scale tracer of LECRIR within the interstellar medium.

At a distance of approximately 400 pc \citep{2019ApJ...879..125Z}, the Orion Molecular Cloud (OMC) complex serves as an ideal laboratory for resolving the spatial distribution of the LECRIR. The region’s proximity and well-characterized environment—supported by extensive legacy datasets in \hi\ and diverse molecular tracers \citep[e.g.,][]{2013MNRAS.431.1296R, 2015ApJS..216...18N, 2021A&A...645A..26R, 2022SCPMA..6599511J}—permit a robust cross-correlation of HINSA with other physical parameters. Leveraging these multi-wavelength constraints, we can perform a comprehensive analysis of the ISM properties across the OMC, providing the necessary resolution to map LECRIR variations on cloud scales.

This paper is structured as follows. In Section \ref{sec:data}, we describe the multi-wavelength datasets and observations utilized in this work. Section \ref{sec:analysis} details our methodology for deriving HINSA column densities, the LECRIR, and local star formation rates (SFRs). Our main results, including the spatial distribution of HINSA, LECRIR and a comparison with \textit{Fermi}-LAT observations, are presented in Section \ref{sec:results}. We discuss the impact of ultraviolet (UV) photodissociation and quantify the primary sources of uncertainty in Section \ref{sec:discussion}. Finally, a summary of our conclusions is provided in Section \ref{sec:summary}.

\section{Data}
\label{sec:data}

\subsection{ \hi\ Data}

The Commensal Radio Astronomy FAST Survey (CRAFTS) \citep{2018IMMag..19..112L} is a key program conducted by the Five-hundred-meter Aperture Spherical Telescope (FAST) \citep{2011IJMPD..20..989N}. This survey aims to simultaneously observe galactic \hi, extra-galactic \hi, pulsars, and Fast Radio Bursts (FRBs) during drifting observations. In March 2023, the CRAFTS group released the first Galactic \hi\ data cube, which covers an area of 600 deg$^2$ \citep{craftsdata}. The declination range of this data cube spans from approximately -11.3\deg\ to 2.3\deg. The final \hi\ data cube has a spatial resolution of approximately 3\arcmin\ (Full Width at Half Maximum; FWHM) and a spectral resolution of 0.2 km s$^{-1}$. The root mean square (rms) value of the noise in each spectrum is approximately 0.17 K (in $T\rm_A$) in channels of velocity width equal to 0.2 \kms. Using the CRAFTS data cube, we generated an \hi\ cube of the Orion region using the \textit{Montage} software\footnote{\url{http://montage.ipac.caltech.edu/}} and show the result in Fig.~\ref{fig:data_map}(a). 

\subsection{CO Data}

The \co\ and \13co\ J=1-0 data of the Orion molecular clouds were acquired using the SEQUOIA focal plane array on the 14m telescope at the Five College Radio Astronomy Observatory (FCRAO) \citep{2013MNRAS.431.1296R}. The rms values for the \co(1-0) data were determined to be 2.0 K, while for the \13co(1-0) data, the rms values were found to be 0.77 K, with a velocity resolution of 0.2 \kms. The \co\ and \13co\ cubes were smoothed to a beam size of 3\arcmin\ and then resampled with a pixel size of 1.5\arcmin. The sky coverage of CO data is shown in Fig.~\ref{fig:data_map}(b). 

\subsection{Extinction Data and Young Stellar Objects Catalog}

We utilized the archival column density of \h2 ($N\rm_{H_2}$) from the \textit{Herschel} Gould Belt Survey \citep{2010A&A...518L.102A, 2013ApJ...763...55R}, which has a spatial resolution of 36.4\arcsec. Following  \cite{2014A&A...562A.138R}, we convert $N\rm_{H_2}$ to visual extinction $A\rm_V$  using a two-step prescription: at low column densities ($N\rm_{H_2} \lesssim 6\times 10^{21}$ \cc), we adopt $A_V/N_{H_2}=1.06\times 10^{-21}$ \cc \citep{1978ApJ...224..132B}; for higher column densities, we adopt $A_V/N_{H_2}=1.45\times 10^{-21}$ \cc \citep{2003ARA&A..41..241D, 2009ApJS..181..321E}. As with the CO data, the extinction map was smoothed to a beam size of 3\arcmin\ and subsequently resampled to a pixel size of 1.5\arcmin.

\begin{figure*}
\begin{center}  
  \includegraphics[ width=0.95\textwidth]{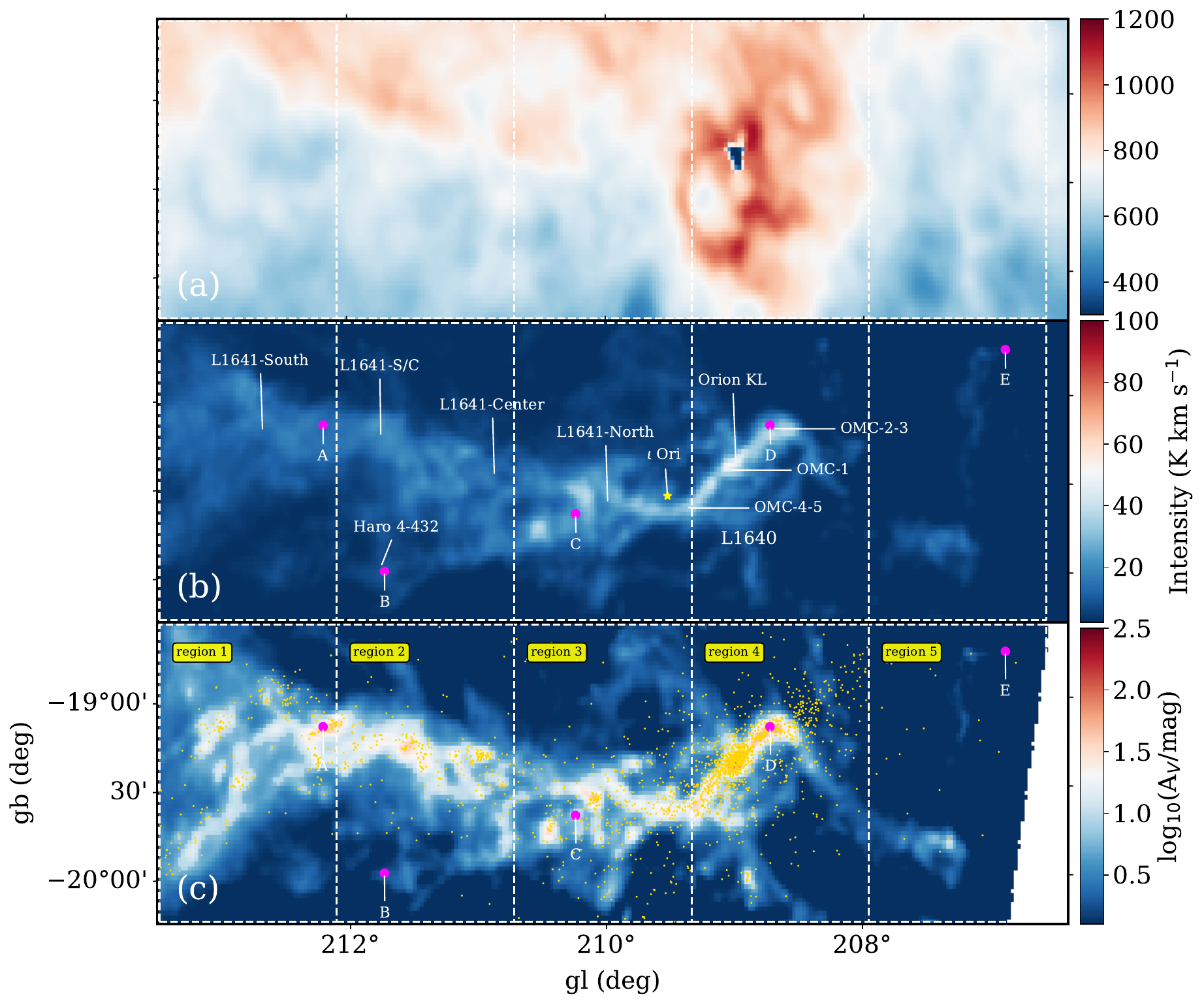}
\caption{ Integrated intensity of \hi\ (Panel a) and \13co(1-0)\ (Panel b) toward the Orion region.  The velocity range is from 0 to 15 \kms\ and a correction for the main beam efficiency has been adopted. The positions of well known clouds and stars are shown in Panel b. The distribution of young stellar objects overlaid on the $Herschel$ extinction map is displayed in Panel c. The boundaries of the five subregions are delineated by white dashed lines,  and the five positions A–-E covering regions with different environments are indicated by magenta dots. Yellow dots indicate the positions of young stellar objects.} 
\label{fig:data_map} 
\end{center}
\end{figure*}

The Young Stellar Objects (YSOs) census toward the Orion region in our analysis primarily relies on the Spitzer Space Telescope survey  \citep{2012AJ....144..192M, 2016AJ....151....5M} and deep near-infrared  VISTA survey \citep{2019A&A...622A.149G}. The distribution of YSOs overlaid on a map of the extinction is shown in Fig.~\ref{fig:data_map}(c).

\subsection{Fermi Data}
\label{sec:fermi_dust}
We collected the latest \textit{Fermi}-{\rm LAT} Pass 8 data from August 4, 2008 (MET 239557417) until May 5, 2025 (MET 768172085), and used the Fermitools from Conda distribution\footnote{\url{https://github.com/fermi-lat/Fermitools-conda/}} for the data analysis.
Due to the extended nature of Orion molecular cloud, we chose the 20\deg\ $\times$ 20\deg\ square region centered at position (R.A. = 84.178\deg, Dec. = -6.304\deg) to be our region of interest (ROI).

We employed the \texttt{gtselect} tool to select events categorized as "source" (with \texttt{evtype = 3} and \texttt{evclass = 128}), restricting to zenith angles $< 90^{\circ}$ to reduce background contamination from the Earth's limb. To select the good time intervals, we applied the recommended expression $\rm (DATA\_QUAL > 0) \&\& (LAT\_CONFIG == 1)$. We also applied the instrument response functions {\it P8R3\_SOURCE\_V3} to analyze the SOURCE events.

\section{Analysis}
\label{sec:analysis}

\subsection{Derivation of N(HINSA)}
\label{subsec:hinsa_calculation}

\begin{figure*}
\begin{center}  
  \includegraphics[width=0.98\textwidth]{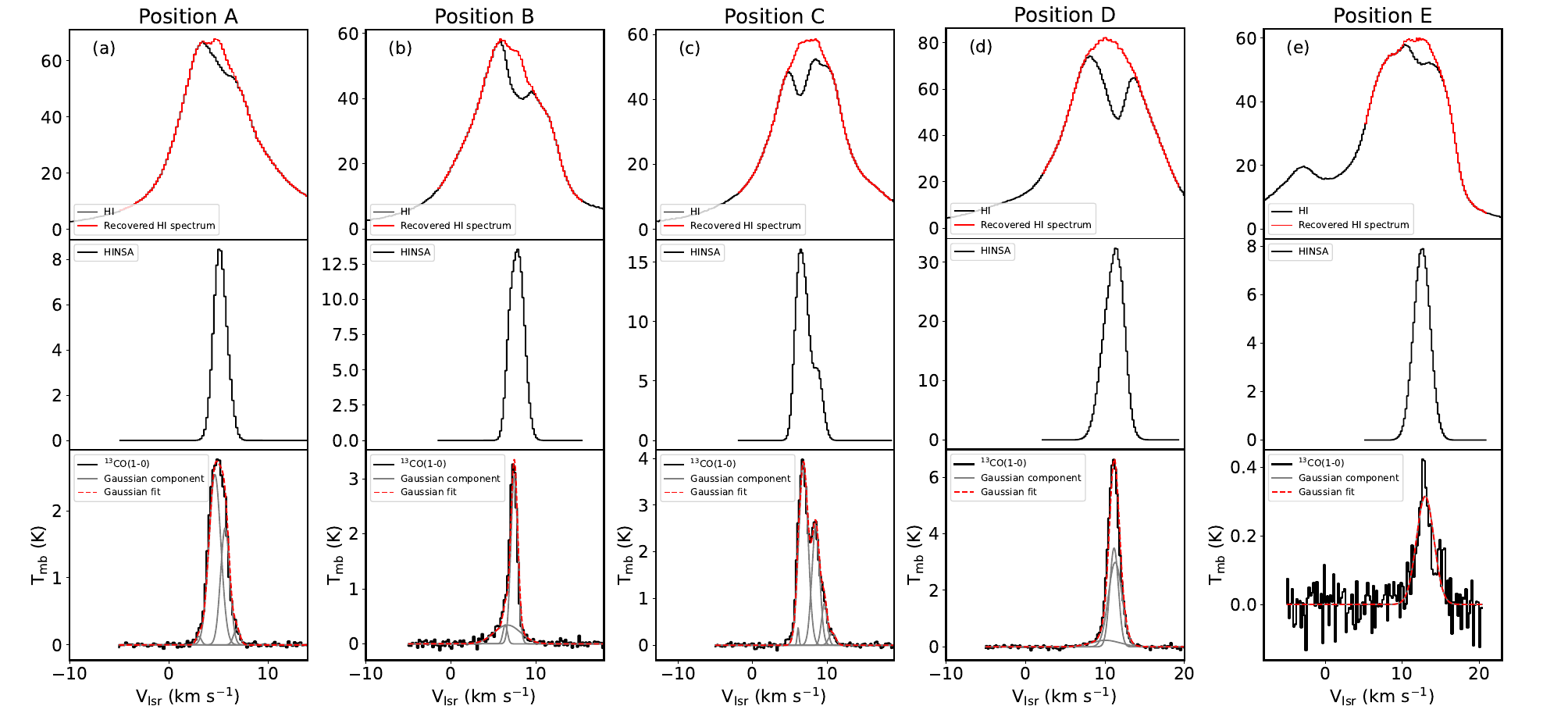}
\caption{Fitting profile of HINSA toward the positions A--E shown in Fig. \ref{fig:data_map}. The top panel shows the observed \hi\ spectrum (black) and the recovered \hi\ spectrum without self-absorption (red). The middle panel displays the HINSA profile. The $^{13}$CO(1–0) spectrum, Gaussian fitting components, and the overall fitting profile are shown as solid black, solid gray, and red dashed lines, respectively. }
\label{fig:hinsa_fit} 
\end{center}
\end{figure*}

To decompose the \13co\ data, we employed the GAUSSPY+ algorithm \citep{2019A&A...628A..78R}. Tests conducted on \13co\ data from the FCRAO telescope demonstrated the capability of the GAUSSPY+ algorithm to handle complex emission with low signal-to-noise ratios. An example of decomposed spectra of the \13co\ data after applying GAUSSPY+ algorithm is presented in Fig.~\ref{fig:hinsa_fit}. By combining the second derivative of the \hi\ spectrum and physical information from CO data, we adopted the effective method for deriving the HINSA column density \citep{2008ApJ...689..276K}. This method requires a high signal-to-noise ratio of the \hi\ spectrum to minimize noise in the second derivative spectrum. The sensitivity of the CRAFTS \hi\ data is inadequate to maximize the extraction of the HINSA signal. Improving the signal-to-noise ratio of the observed \hi\ profile by fitting a multi-order polynomial \citep{2022A&A...658A.140L}  is not suitable for this study as the noise rms in the CRAFTS data is approximately 6-10 times higher.

To mitigate noise amplification during the numerical derivative process, we followed the approach in \cite{https://doi.org/10.5402/2011/164564} and introduced total-variation regularization. The value of the regularization parameter, $\alpha=0.2$, was selected. Our tests indicate a weak dependence on the $\alpha$ value. The absorption profile, $T_{ab}(v)$ is  described by the equation \citep{2003ApJ...585..823L}:
\begin{equation}
    T_{ab}(v)=[pT_{HI}(v)+(T_c-T_{ex})(1-\tau_f)](1-e^{-\tau}),
\end{equation}
where $T_c$ and $T_{ex}$ are the background continuum temperature and the excitation temperature, respectively. The value for $T_c$ is derived from the CHIPASS continuum survey at 1.4 GHz \citep{2014PASA...31....7C}.  Since the  CHIPASS data is in unit of full-beam temperature $T\rm_{fb}$, it was converted into main-beam brightness temperature with T$_c$=$(T_{fb}-2.73)/0.6$ +2.73 \citep{2022MNRAS.512.3345D}. The term $\tau_f$ represents the optical depth of the foreground \hi\ gas.  The parameter $p$ is defined as the fraction of $\tau_f$  relative to the total \hi\ optical depth along the line of sight. Constant values of $\tau_f=0.1$ and $p=0.9$ were adopted for calculations. Fig.~\ref{fig:hinsa_fit} presents an example of derived HINSA profile.

The HINSA column density N(HINSA) is calculated using the following equation \citep{2003ApJ...585..823L}:  

\begin{equation}
\rm N(HINSA) = 1.9\times 10^{18}\tau T_{ex}\Delta V \rm\ cm^{-2}, 
\end{equation}
where $\tau$, $T_{ex}$, and $\Delta V$ are the optical depth, excitation temperature, and FWHM of the HINSA emission, respectively.  The excitation temperature ($T\rm_{ex}$) of HINSA, a key parameter for deriving N(HINSA), is adopted from the $^{12}$CO(1-0) line through $T_{ex} = 5.5/ln(1+5.5/(T_{mb}(^{12}CO)+0.82))$, in which $T\rm_{mb}(^{12}CO)$ represents the peak brightness temperature of \co(1-0). This is valid for our study, which targets dense regions ($A\rm_V> 5$ mag) of the Orion molecular cloud where the gas density exceeds the critical density of $\sim 10^2$ cm$^{-3}$ for \co(1-0) transition in region with optical depth of $\sim$ 20, thus ensuring local thermodynamic equilibrium.

\subsection{Low-energy Cosmic Ray Ionization Rate}
\label{subsec:lecr}

By neglecting the effects of UV photo-dissociation, the time-dependent abundance of HINSA, $x(t)$, at the center of a molecular cloud, arising from ionization by Low-Energy Cosmic Rays, is given by \citep{2005ApJ...622..938G}: 
\begin{equation}
x(t) = 1-\frac{2k'n_0}{2k'n_0 + \zeta}[1-exp(\frac{-t}{\tau_{c}})], 
\label{eq:hinsa_x}
\end{equation}
where the parameters $k'$, $n_0$ and $\zeta$ represent  the formation rate coefficient of \h2\ , the total proton density,  and the cosmic-ray ionization rate, respectively. The composition and size distribution of grains can strongly affect the value of $k'$ by an order of magnitude \citep{2004A&A...414..531H, 2005ApJ...622..938G}. Assuming a common grain composition and size distribution, the formation rate coefficient of \h2\ varies as a function of temperature, with $k' = 1.2 \times 10^{-17} \sqrt{T_k/10\ K}\ \text{cm}^3\ \text{s}^{-1}$  \citep{2005ApJ...622..938G}. \citet{2011piim.book.....D} suggests a consistent value of $k' \approx 3 \times 10^{-17} \sqrt{T_k/70\ K}\ \text{cm}^3\ \text{s}^{-1}$. However, based on UV absorption of \h2,  \citet{2008ApJ...680..384W}  found a lower rate of $k' = 1 \times 10^{-17}\ \text{cm}^3\ \text{s}^{-1}$ toward diffuse regions ($A_V \lesssim 0.25$ mag) and $k' \sim 3.5 \times 10^{-17}\ \text{cm}^3\ \text{s}^{-1}$ toward denser regions ($0.25 \leq A_V \leq 2.13$ mag). This indicates that shock processing may decrease the grain surface area in diffuse regions. Nevertheless, we adopted the general expression from \citet{2005ApJ...622..938G}, which gives $k' = 1.2 \times 10^{-17}\ \text{cm}^3\ \text{s}^{-1}$ at $T_k = 10$ K.

To calculate total proton density,  we adopted the relationship between the averaged number density of hydrogen, $n_0 = n$(\hi) + 2$n$(\h2), and the observed visual extinction, $A_{V,obs}$ \citep{2023MNRAS.519..729B}:

\begin{equation}
    <n_0> = (\frac{A_\mathrm{V,obs}}{0.06})^{1/0.69}\rm cm^{-3}. 
\label{eq:n0_AV}
\end{equation}
This equation is based on various hydrodynamic simulations spanning scales from parsecs to kiloparsecs \citep{2010MNRAS.404....2G, 2013ApJ...764...36V, 2017MNRAS.465..885S, 2017MNRAS.472.4797S, 2021ApJ...920...44H}  and reflects the most probable relationship in well-shielded clouds with solar metallicity (see also \citet{2025A&A...700L..16G} for an analytical approach).  We applied this equation to calculate $n_0$ in the Orion complex. 

The derived average density $<n_0>$ toward the OMC-2 region is $\sim 2 \times 10^4$ cm$^{-3}$, corresponding to an \h2\ column density N(\h2) of $3.5\times 10^{22}$ cm$^{-2}$. For comparison, N$_2$H$^+$ observations yield a proton density ($n_H=2n_{H_2}$) of $2\times 10^5$ cm$^{-3}$ within OMC-2 cores, corresponding to N(\h2) of $\sim 1\times 10^{23}$ cm$^{-2}$ \citep{2008PASJ...60..407T}. One factor contributing to this difference is the lower N(\h2) value resulting from beam dilution (beam size of 3 arcmin in this study), which reduces the inferred density by a factor of five. 
Another source of uncertainty may arise from the estimation of the core radius $R$ in \citet{2008PASJ...60..407T}, which directly affects the derived core density through $n(H_2)=N(H_2)/(2R)$.

The parameter $t$ denotes the evolutionary time, while $\tau_c$ represents the characteristic timescale for the transition from \hi\ to \h2. In dense regions, $\tau_c$ can be approximated as

\begin{equation}
    \tau_c = \frac{2.6\times10^9}{n_0}\rm yr. 
\end{equation}

For a molecular cloud with a number density of $10^4$ \cc, the transition timescale is $\tau_c = 2.6\times10^5$ yr. In the limit of $t \gg \tau_c$, the system reaches a steady state, and the low-energy cosmic-ray ionization rate $\zeta$ (LECRIR) can be expressed as

\begin{equation}
   \zeta = \frac{2k'n_0 x(t)}{1-x(t)}. 
\label{eq:zeta_h2}
\end{equation}

We emphasize that the steady-state assumption provides an upper limit on the LECRIR, with the related uncertainties discussed in detail in Section~\ref{sec:uncertainty}. 

\subsection{Calculation of the Star Formation Rate}
\label{subsec:SFR}

The star formation rate (SFR) serves as a fundamental diagnostic parameter for quantifying star formation activity within molecular clouds. In this study, we employ young stellar objects (YSOs) as tracers to estimate the SFR through the relation:

\begin{equation}
 \mathrm{SFR} = \frac{N_\mathrm{YSO}\cdot \overline{M}_\mathrm{IMF}}{t_\mathrm{YSO}},
\end{equation}
where $N_{YSO}$ represents the number of YSOs, $\overline{M}_{IMF}$ denotes the mean stellar mass derived from the initial mass function (IMF), and $t_\mathrm{YSO}$ corresponds to the average YSO lifetime. Adopting well-established values from previous studies \citep{2010ApJ...724..687L,2009ApJS..181..321E, 2014ApJ...782..114E}, we assume $\overline{M}_\mathrm{IMF}=0.5 M_{\odot}$ and $t_\mathrm{YSO}\approx 2$ Myr for the Orion molecular cloud complex.  This yields the practical expression:

\begin{equation}
    \mathrm{SFR} = 0.25 N_\textrm{YSO}\times 10^{-6} M_{\odot}\rm yr^{-1}. 
\end{equation}

The YSO census toward the Orion region in our analysis primarily relies on the Spitzer Space Telescope survey  \citep{2012AJ....144..192M, 2016AJ....151....5M} and deep near-infrared  VISTA survey \citep{2019A&A...622A.149G}. 
In this study, we adopt four classes of YSOs, including Class 0/I protostars, Flat-spectrum sources, and Class II/III pre-main-sequence stars with disks, all of which are considered capable of driving protostellar jets.

\section{Results}
\label{sec:results}

\subsection{Spatial distribution of N(HINSA)}
\label{subsec:NHINSA}

\begin{figure*}
\begin{center}  
  \includegraphics[width=0.95\textwidth]{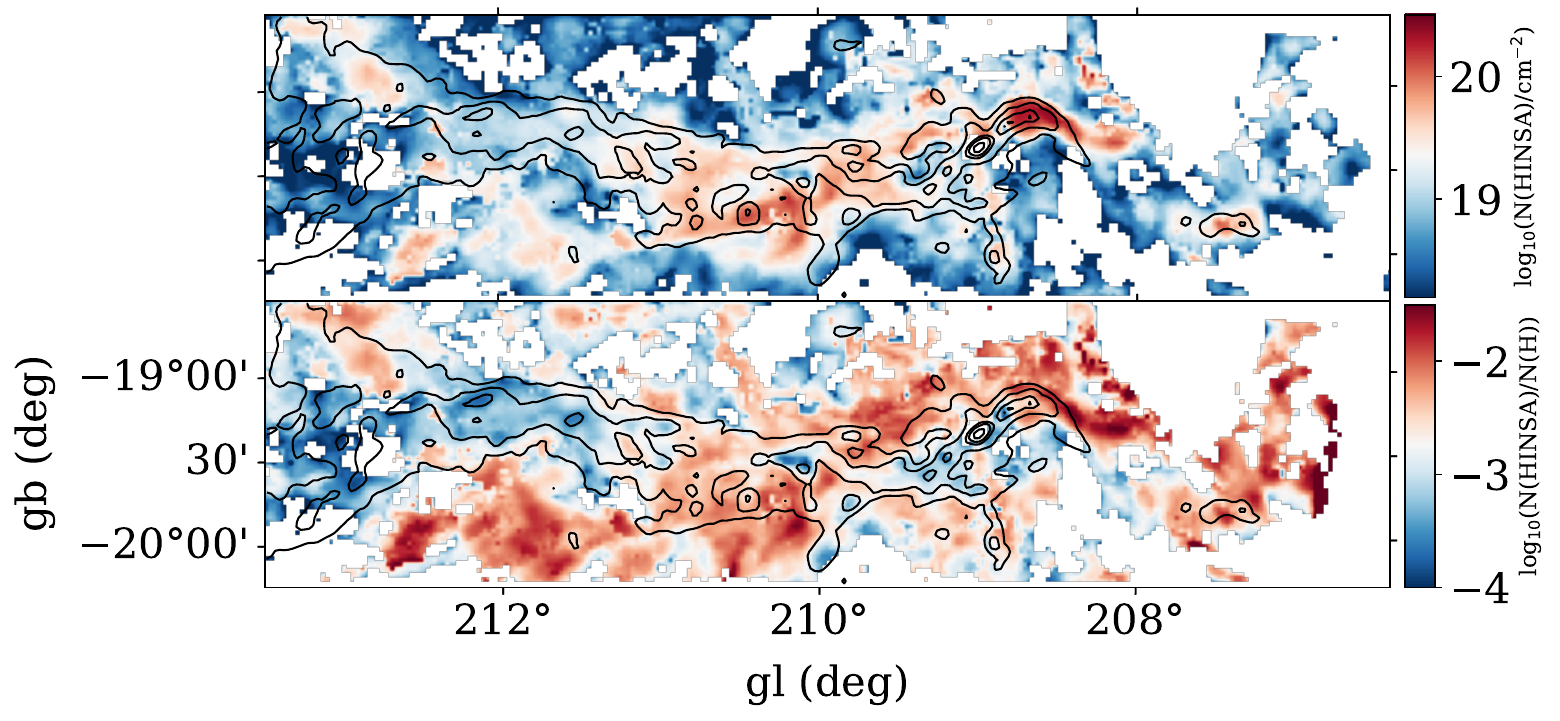}
\caption{Spatial distribution of the HINSA column density, $N$(HINSA) (top panel), and the HINSA abundance relative to total proton column density (bottom panel). The black contours indicate visual extinction of $A\rm_V$= [3, 10, 30, 50, 100] mag. } 
\label{fig:hinsa_map} 
\end{center}
\end{figure*}

The HINSA column density and HINSA abundance ($x$(HINSA)=$N$(HINSA)/(2$N$(\h2))) map toward the Orion complex is presented in Fig. \ref{fig:hinsa_map}. $N(\text{HINSA})$ ranges from $6.5\times 10^{16}$ cm$^{-2}$ to $3.6\times 10^{20}$ cm$^{-2}$ with a median value of $8.5\times 10^{18}$ cm$^{-2}$. The corresponding HINSA abundance value ranges from $2.1\times 10^{-5}$ to $7.4\times 10^{-2}$ with a median value of $2.2\times 10^{-3}$.

By comparing the HINSA distribution with the visual extinction map, we find an overall good spatial correspondence between the two, except in the subregions L1641-S/C and L1641-South, where the HINSA column density and abundance are significantly reduced relative to the surrounding areas.  Hydrodynamical simulations \citep[e.g.,][]{2007A&A...465..431H, 2015MNRAS.452.1353H} predict that cold \hi\ is preferentially distributed along the periphery of dense molecular gas as the cloud evolves from atomic to molecular form. Observationally, \citet{2018ApJ...867...13Z} identified an ``onion-like’’ HINSA shell surrounding the dense core of B227, which was interpreted as evidence for an ongoing \hi-to-\h2\ transition.

The reduced HINSA column densities observed toward L1641-S/C and L1641-South may indicate that these subregions are in a transitional evolutionary stage, where atomic hydrogen is being rapidly converted into molecular hydrogen. If this interpretation is correct, the local gas may not yet have reached chemical steady state. In such a situation, the steady-state approximation adopted in Equation~(\ref{eq:zeta_h2}) would lead to an overestimation of the low-energy cosmic-ray ionization rate, since the residual \hi\ abundance still retains memory of the cloud’s time-dependent chemical evolution. Therefore, the LECRIR values derived under the steady-state assumption should be regarded as upper limits in these chemically young regions. 

\subsection{Relationship of LECRIR with both Extinction and Star formation rate}
\label{subsec:cr_vs_AV}

\begin{figure*}
\begin{center}  
  \includegraphics[ width=0.95\textwidth]{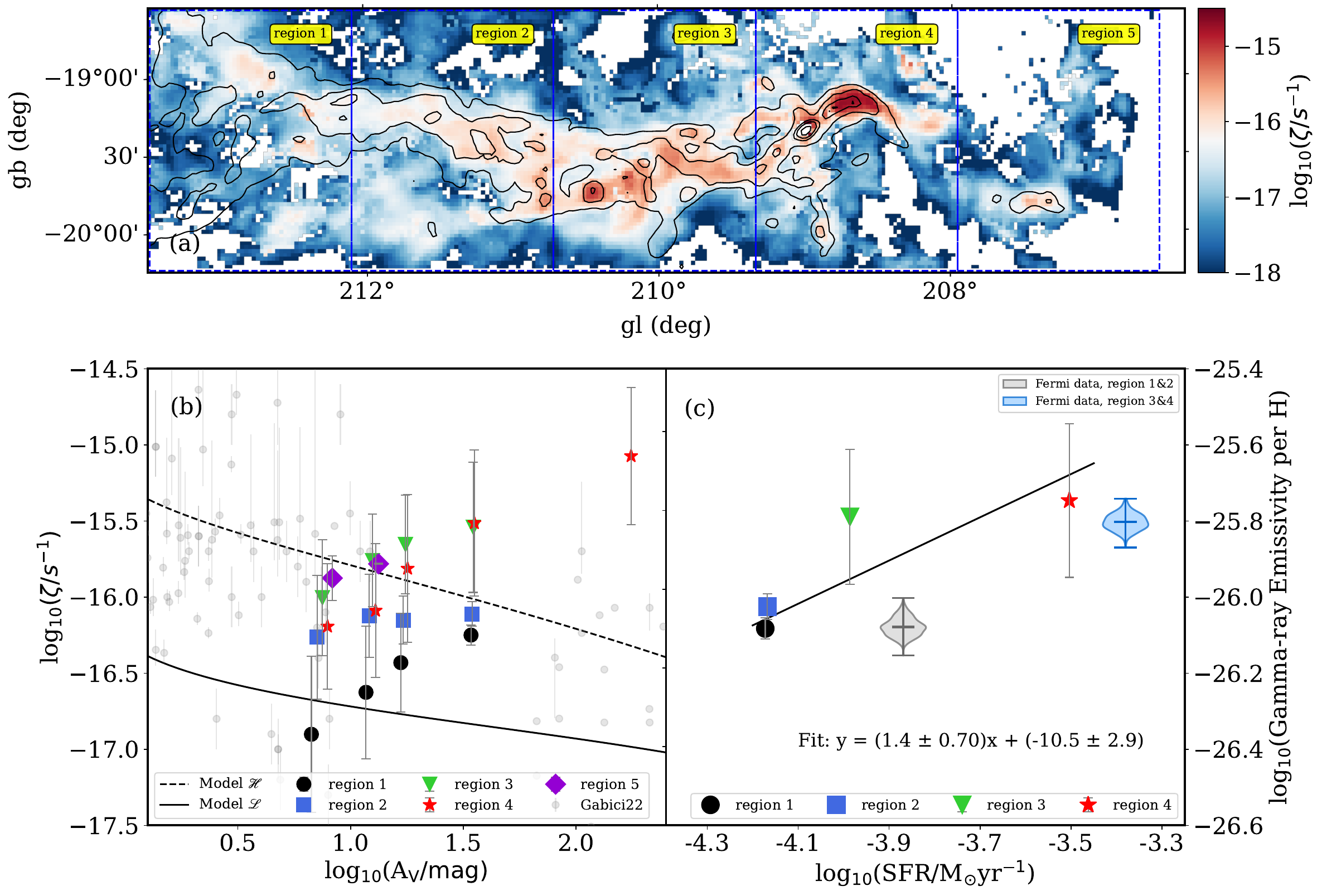}
\caption{ (a) Spatial distribution of the low-energy cosmic ray ionization rate (LECRIR); (b)  Median LECRIR values within five visual-extinction ranges ([5–10], [10–15], [15–20], [20–50], and [50–300] mag) across the five subregions of the Orion molecular complex. The representative $A_{\rm V}$ values for these ranges are 7.5, 12.5, 17.5, 35, and 175 mag, respectively. Small offsets have been applied to the $A_{\rm V}$ values of each subregion for clarity. Also shown are the $Voyager$-based $\mathcal{L}$ model (solid black line) and the reproduced diffuse-region $\mathcal{H}$ model (dashed black line) \citep{2018A&A...619A.144P, 2024A&A...682A.131P}, along with previous LECRIR measurements (gray dots) derived from ion tracers such as H$_3^+$, OH$^+$, and HCO$^+$ \citep{2022A&ARv..30....4G, 2023ApJ...950...64I}.;  (c) Relationship between LECRIR and star formation rate for 4 subregions. The LECRIR for each subregion is estimated using the median value derived from areas with $A_{\rm V} \geq 15$ mag. The linear fit between LECRIR and star formation rate is shown with black solid line. The Gamma-ray emissivities per hydrogen from the $Fermi$ survey for the two groups (Regions 1 \& 2 and Regions 3 \& 4) are shown as violin plots  for comparison, with their scale on the right hand axis.}
\label{fig:lecrir_map} 
\end{center}
\end{figure*}

By combining the derived HINSA abundance with the volume density estimated from the visual extinction A$\rm_V$ along each line of sight, we constructed the LECRIR map shown in Figure~\ref{fig:lecrir_map}(a). The inferred LECRIR values across the Orion region span nearly three orders of magnitude, ranging from $\sim 1\times10^{-18}$ to $5\times10^{-15}\ {\rm s^{-1}}$. 

For the Horsehead Nebula, we derive a LECRIR value of $\sim 5\times10^{-16}\ {\rm s^{-1}}$, consistent with previous estimates by \citet{2012A&A...537A...7R}. In contrast, toward OMC-2 FIR 4, our measured value of $\sim 8\times10^{-16}\ {\rm s^{-1}}$ is substantially lower than the $\sim 4\times10^{-14}\ {\rm s^{-1}}$ reported in earlier studies \citep{2014ApJ...790L...1C, 2017A&A...605A..57F, 2018ApJ...859..136F}. This discrepancy may arise from at least two factors. First, beam dilution is likely significant: the compact OMC-2 FIR 4 region has a characteristic size of only $\sim$ 30\arcsec, whereas our effective beam size is $\sim$ 180\arcsec. Second, previous determinations rely heavily on assumed abundances of molecular ions (e.g., c-C$_3$H$_2$, with abundances on the order of $10^{-12}$ and on detailed chemical modeling, both of which introduce substantial systematic uncertainties.

Fig. \ref{fig:lecrir_map}(b) presents the relationship between LECRIR and visual extinction $A\rm_V$. 
For comparison, we include two models for the ionization rate in molecular clouds illuminated by external LECRs: model $\mathcal{L}$, which incorporates data from the two Voyager spacecraft \citep{2016ApJ...831...18C, 2019NatAs...3.1013S} and features a low-energy spectral slope, and model $\mathcal{H}$, which reproduces the average LECRIR value in diffuse regions \citep{2018A&A...619A.144P, 2024A&A...682A.131P}. The relationship between $\zeta$ and $A_V$ exhibits significant variation across five distinct sub-regions (denoted as Region 1 to Region 5), particularly in areas where $A_V > 5 \text{ mag}$. In each of these regions, $\zeta$ is observed to increase with extinction, rising from $A_V \approx 5 \text{ mag}$ to $A_V \approx 175 \text{ mag}$. The values of $\zeta$ in the high-extinction zones of Regions 3 and 4 are comparable to or even exceed the average low-energy cosmic-ray ionization rate (LECRIR) from model $\mathcal{H}$.  These observations suggest that the low-energy cosmic rays detected in dense regions are unlikely to originate from the propagation of particles from the less-shielded outer environment. Furthermore, this trend is consistent with chemical models predicting that the cosmic-ray ionization rate increases with visual extinction in regions containing embedded protostars \citep{2019ApJ...878..105G}.

A series of observations of ions  conducted in star-forming regions indicate higher ionization rates; however, these findings rely on model-dependent measurements of ions \citep{2018ApJ...865..127I, 2024A&A...690A.293L, 2024A&A...686A.162P}.  HINSA, a direct product of H$_2$ dissociation by LECRs, is much less dependent on chemical models. HINSA analysis reveals the  trend of increasing LECRIR toward higher  extinction, which is the antithesis of LECRs' propagation effect from outside the cloud and is consistent with in situ production of LECRs.  In situ particle acceleration --- driven by shocks along protostellar jets or on protostellar surfaces \citep{2015A&A...582L..13P, 2016A&A...590A...8P, 2018ApJ...861...87G}  --- can explain the enhanced  ionization rate observed in the star-forming region OMC-2 FIR 4  \citep{2014ApJ...790L...1C, 2017A&A...605A..57F, 2018ApJ...859..136F} and the Central Molecular Zone \citep{2019Natur.573..235H}.

Figure~\ref{fig:data_map}(c) presents the spatial distribution of YSO candidates in five regions with same area (1.3\deg$\times$1.675\deg) corresponding to 106.5 pc$^2$ (9.10$\times$ 11.7 pc$^2$) at distance of 400 pc. Table~\ref{tab:yso_numbers} summarizes the YSO populations, SFR, and SFR surface density in the five sub-regions. The star formation activity in region 3 and region 4 is significantly more intense compared to those in region 1 and region 2 \citep{2010ApJ...724..687L, 2019A&A...622A.149G}. 
As shown in Figure \ref{fig:lecrir_map}(c), the LECRIR value for $A_V> 10$ mag is proportional to star formation rate (SFR) following:

\begin{equation}
log_{10}\zeta = (1.4\pm 0.70)log_{10}\mathrm{SFR} + (-10.5\pm 2.9). 
\end{equation}

\begin{table}
\centering
\caption{Number of YSOs and SFR in the five regions of Orion molecular cloud. }
\begin{tabular}{lccccc}
\hline \hline  
Sub-region & Number & SFR [$10^{-4} M_{\odot}\ \rm yr^{-1}$] \\
\hline 
region 1 & 269 & 6.72$\times 10^{-1}$   \\ 
region 2 & 272 & 6.80$\times 10^{-1}$   \\ 
region 3 & 413 & 1.03   \\ 
region 4 & 1254 & 3.14   \\ 
region 5 & 23 &  5.75$\times 10^{-2}$   \\ 
\hline 
\end{tabular}
\label{tab:yso_numbers}  
\end{table}

This positive correlation suggests that the cosmic-ray population in these dense regions is not dominated by the penetration of the broader Galactic background. Instead, it provides strong evidence that low-energy cosmic rays are produced \textit{in situ} by local star-forming processes, which act as the primary drivers for the elevated ionization rates observed in the OMC's most active subregions.

\subsection{Support from \textit{Fermi}-LAT Gamma-ray Observations}
\label{subsec:fermi-results}

For cosmic rays more energetic than 280 MeV, $\gamma$-rays can be generated through the p-p (proton-proton) interaction, thus offering an independent constraint of LECR \citep{2014pp}. 
 Indeed, the gamma-ray observations reveal significant smaller gamma-ray emissivities below several GeV in dense clumps inside GMCs, which also indicate that the Galactic cosmic rays are effectively shielded and can hardly penetrate into the high density regions \citep{yang23}. 
 Moreover, the recent gamma-ray observations also reveal particle acceleration in protostellar systems \citep{yan22}.  
 
\begin{figure*}
\begin{center}  
  \includegraphics[width=0.6\textwidth]{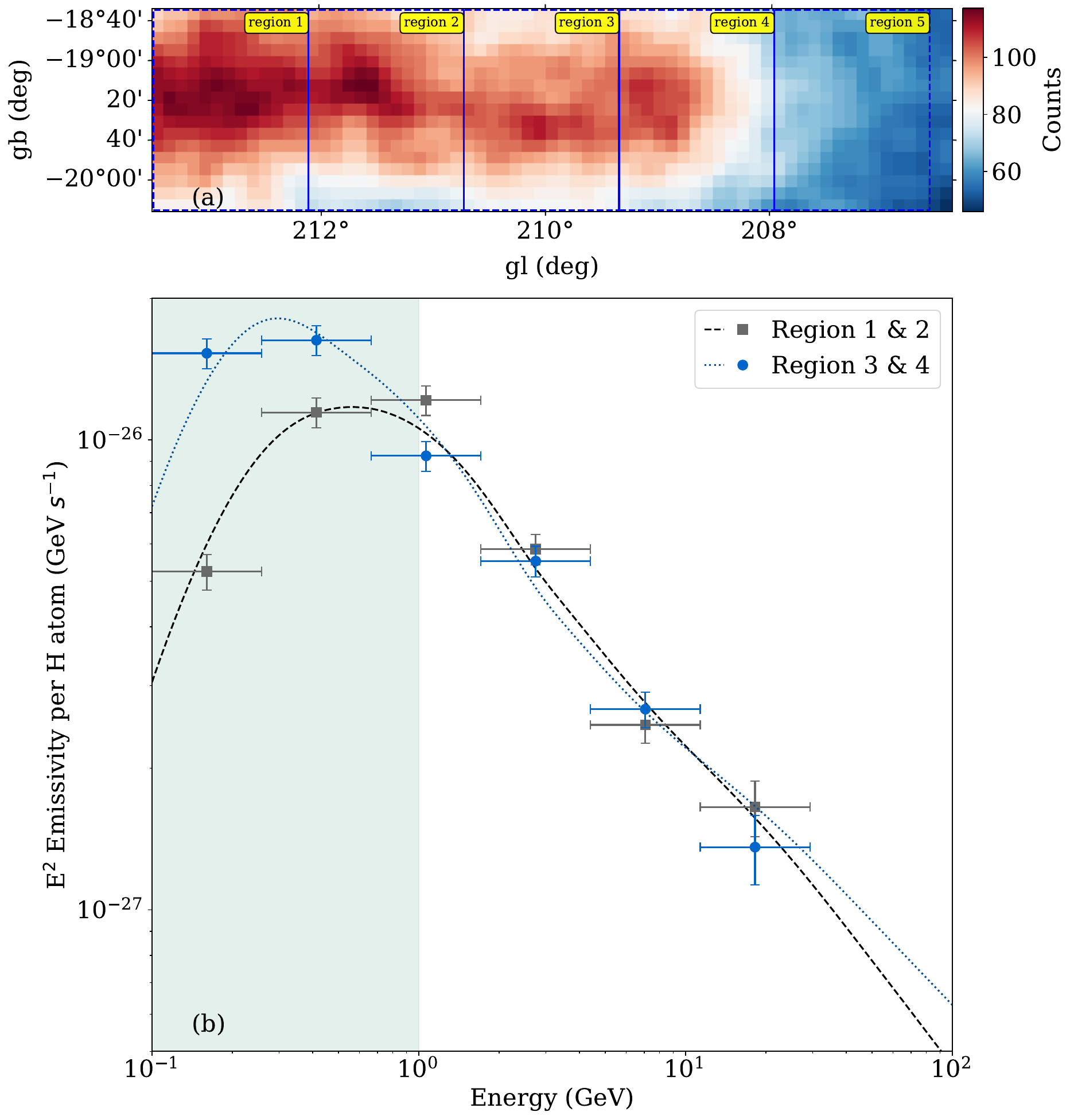}
\caption{ (a) Gamma-ray counts map across the energy range of 0.1 to 1 GeV; (b) The derived gamma-ray emissivity per hydrogen atom from different regions of Orion molecular cloud. The light green region represents the energy range from 0.1 to 1 GeV. }
\label{fig:fermiemi} 
\end{center}
\end{figure*}

 Utilizing the 0.1 to 1 GeV data from the \textit{Fermi}-LAT \citep{2020ApJS..247...33A}, we derived the gamma-ray counts in the five sub-regions, as shown in Fig. \ref{fig:fermiemi}(a). Due to low photon counts and the poor angular resolution in lower energy bands, we combined Region 1 and 2, as well as Region 3 and 4, in order to derive the $\gamma$-ray emissivity per hydrogen atom, which is proportional to LECR rate \citep{2014pp}, in conjunction with Planck dust data \citep{2014Planck}.
 In the two lowest energy bins, the emissivity of Region 3 \& 4 is estimated to be approximately 1.5 to 2.5 times higher than that of Region 1 \& 2 (see Fig.~\ref{fig:fermiemi}(b)). This finding aligns well with our LECRIR measurements based on HINSA (see Fig.~\ref{fig:lecrir_map}(c)), affirming the star formation's role in cosmic ray production.

\section{Discussion}
\label{sec:discussion}

The spatial distribution of LECRIR across various regions is influenced by both the uncertainties inherent in our calculations and the physical mechanisms governing the transport of LECRs. We delve into these factors in the following subsections.

\subsection{Uncertainties}
\label{sec:uncertainty} 

Our study is built upon three primary assumptions: the density-extinction relationship, the steady-state condition of HINSA, and the exclusion of UV radiation effects.

\begin{enumerate}
\item   Extinction--density  relationship. The average number density is related to the visual extinction along the line of sight through Equation~\ref{eq:n0_AV}, assuming arbitrary three-dimensional density distributions and chemical equilibrium in the photodissociation region (PDR) \citep{2023MNRAS.519..729B}. Although it is difficult to quantify the associated uncertainty, this relationship suggests that the uncertainty decreases toward higher $A_{\rm V}$, where the gas dynamics become increasingly gravity dominated. In contrast, regions with lower $A_{\rm V}$ are more strongly influenced by turbulence, leading to a larger dispersion in the extinction--density relation  \citep{2025A&A...700L..16G}. Since our analysis primarily focuses on dense gas, the adopted $A_{\rm V}$--$n_{\rm H}$ relationship should remain reasonably reliable.

\item Steady-state assumption. The chemical dynamics within translucent clouds are not static.  \citet{2007ApJ...654..273G} demonstrate that the HINSA profile is temporally variant, transitioning from a dual-absorption feature to a singular one. Consequently, a steady-state assumption provides an upper limit of the LECRIR value. In order to estimate the uncertainties, we plot the variation of ratio between LECRIR under steady-state and LECRIR in different chemical ages, $t=1$ Myr and $t=7$ Myr. As shown in Fig. \ref{fig:zeta_ratio_time}, the ratio ranges from 1 to $\sim 10$, however, over most of the relevant parameter space, the ratio remains below 1.2,  indicating that the steady-state approximation introduces only a modest bias in the majority of cases.

\begin{figure*}
\begin{center}  
  \includegraphics[width=0.48\textwidth]{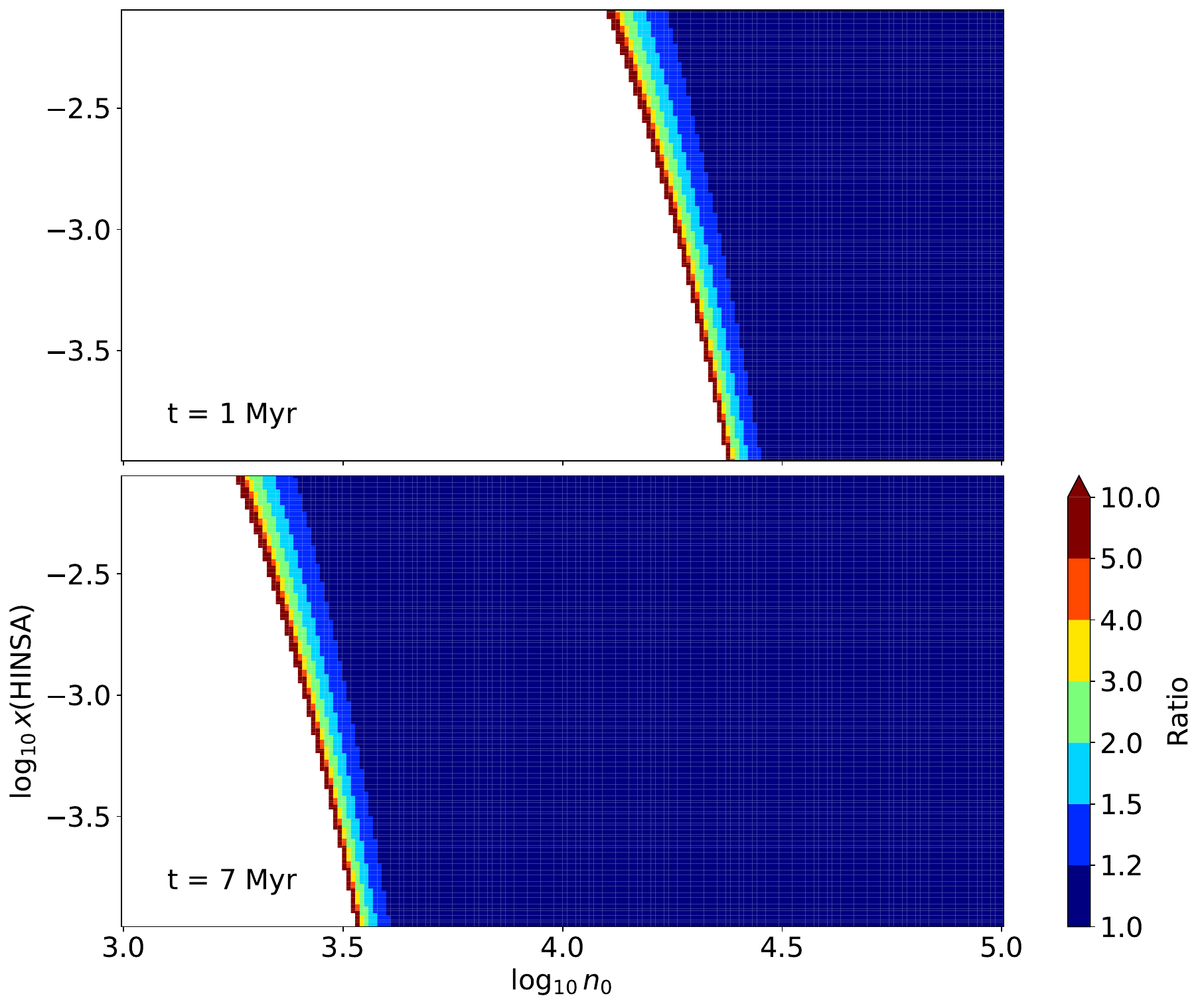}
\caption{ Distribution of the ratio between the LECRIR derived under the steady-state assumption and that derived for finite chemical ages (Equation \ref{eq:hinsa_x}), shown as a function of gas number density $n_0$ and HINSA abundance $x(\mathrm{HINSA})$. The upper and lower panels correspond to chemical ages of $t=1$ Myr and $t=7$ Myr, respectively. White regions indicate NaN values, where the HINSA abundance cannot be reproduced under the steady-state assumption. }
\label{fig:zeta_ratio_time} 
\end{center}
\end{figure*}

\end{enumerate}

In summary, while the assumptions regarding the extinction--density  and the steady state of HINSA may introduce uncertainties in specific locales, they do not affect the overall results in Section \ref{sec:results}.

\subsection{Contribution of HINSA from UV Photodissociation}
\label{subsec:hinsa_uv}

Through photodissociation of \h2 \citep{1988ApJ...334..771V}, UV photons could generate \hi\ too, which may affect the contribution of HINSA by LECRs. 
We investigate the contribution of HINSA from UV and LECRs by performing photodissociation region models using {\sc 3d-pdr} \citep{Bisbas12}. We perform two suites of models; one using a constant density slab ($n_{\rm H}=10^3\,{\rm cm}^{-3}$), and one using the `variable density slab' \citep{Bisbas19,2023MNRAS.519..729B}. The latter function (dubbed as `$A_{\rm V}$-$n_{\rm H}$-relation') is found to best represent the PDR chemistry of complex three-dimensional structures at a minimal computational cost. Each of these density distributions interact with two different intensities of impinging FUV radiation ($G_0=10$ and $G_0=1000$, normalized to the spectral shape of Draine 1978), and two different LECRIR ($\zeta_{\rm CR}=10^{-20}\,{\rm s}^{-1}$ representing the no-cosmic-ray scenario, and $\zeta_{\rm CR}$ following the $\cal H$-function of cosmic-ray attenuation \citep{Padovani18,Gaches22}. In all models, the \h2\ formation follows the description of \citep{Cazaux02,Cazaux04} and we account for the ion-recombination on dust grains (including a population of PAH grains) \citep{Weingartner01}.

\begin{figure*}
\begin{center}  
  \includegraphics[width=0.48\textwidth]{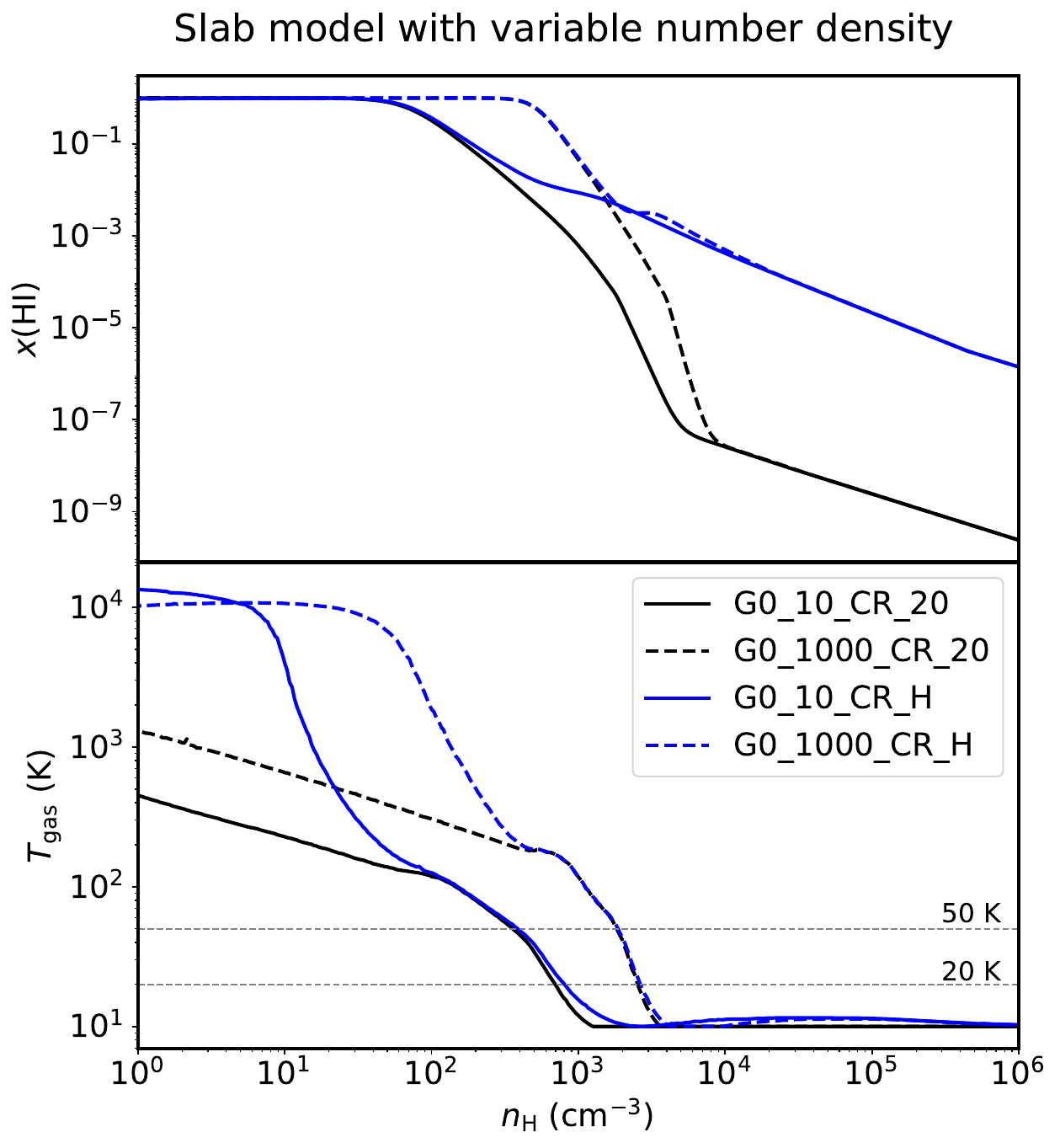}
  \includegraphics[width=0.48\textwidth]{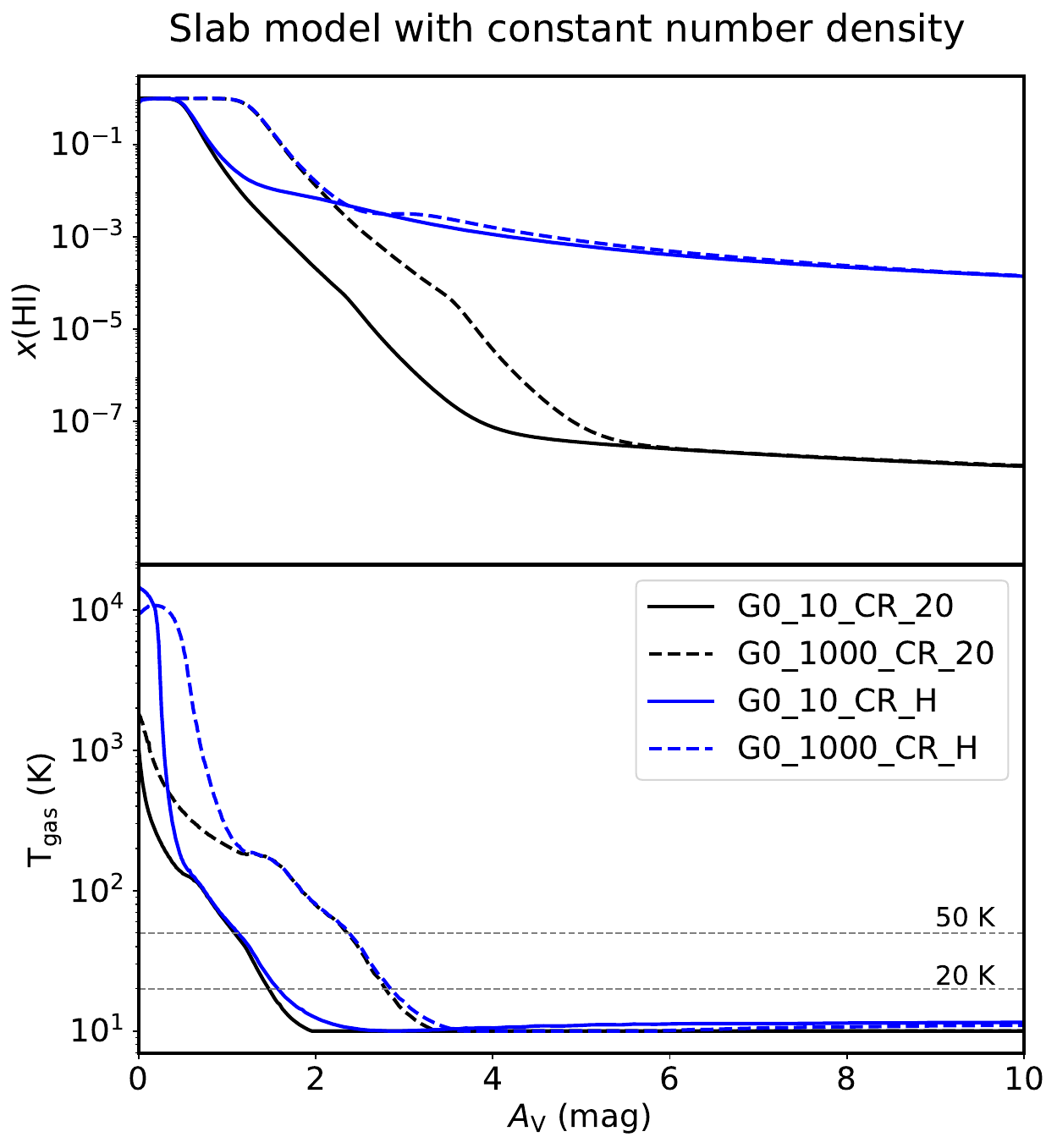}
\caption{Derived H{\sc i} abundance (upper panels) and gas temperatures (lower panels) for a variable density slab (left panels) and a constant density slab (right panels) using the {\sc 3d-pdr} code \citep{Bisbas12}. The left panels are a function of the $n_{\rm H}$ number density, whereas the right panels a function of the visual extinction, $A_{\rm V}$. For each case four models were run: (1) UV intensity $G_0=10$ with LECRIR of $1\times10^{-20}\,{\rm s}^{-1}$ (solid black lines); (2) UV intensity $G_0=10^3$ with LECRIR of $1\times10^{-20}\,{\rm s}^{-1}$ (dashed black lines); (3) UV intensity $G_0=10$ with LECRIR following the $\cal H$ function \citep{Padovani18} (solid blue lines); (4) UV intensity $G_0=10^3$ with LECRIR following the $\cal H$ function \citep{Padovani18} (dashed blue lines). }
\label{fig:pdrmodels} 
\end{center}
\end{figure*}

As shown in Fig. \ref{fig:pdrmodels}, the contribution of LECRs to the \hi\ abundance $x$(\hi)=$n$(\hi)/($n$(\hi)+$2n$(\h2)) exceeds that of UV photons by approximately two orders of magnitude in high-density region (e.g., $n{\rm _H} \gtrsim 3 \times 10^3 \ {\rm cm}^{-3}$ in the variable-density slab model) and in dense regions (e.g., $A_{\rm V} \gtrsim 2.5 \ {\rm mag}$ in the constant-density slab model). Besides, the inclusion of LECRs increase the gas temperature by several Kelvin.

Given that HINSA originates from within dense molecular gas, typically characterized by temperatures of $\sim 20$ K, it is effectively shielded from the interstellar radiation field. We therefore conclude that UV photodissociation exerts a negligible impact on our derived LECRIR, reinforcing the interpretation that LECRs are the primary drivers of cold \hi\ gas formation in dense environments.

\section{Summary}
\label{sec:summary}

This study presents a detailed map of the Low-Energy Cosmic Ray Ionization Rate (LECRIR) across a $\sim$ 12 deg$^2$ region of the Orion Molecular Cloud. Leveraging high-sensitivity \hi\ data from the Commensal Radio Astronomy FAST Survey (CRAFTS), we applied the HINSA method to derive LECRIR values with minimal dependence on complex chemical modeling. Our key findings are as follows:

\begin{enumerate}
    \item HINSA Distribution: We determined a median HINSA column density of $8.5 \times 10^{18}$ cm$^{-2}$. The positive correlation observed between N(HINSA) and the total proton column density N(H) in the central Orion complex is consistent with a steady-state \hi--\h2\ transition within the molecular interior.

    \item  LECRIR and Extinction: The LECRIR ($\zeta$) increases with visual extinction from $A_V \approx 5 \text{ mag}$ to 175 mag. In the high-extinction zones of Regions 3 and 4, $\zeta$ values are comparable to or exceed model-$\mathcal{H}$ \citep{2018A&A...619A.144P} predictions (Fig. \ref{fig:lecrir_map}b). This trend suggests that LECRs in dense cores are not merely remnants of particles penetrating from the diffuse exterior, but are likely generated internally.

    \item  Link to Star Formation: For $A\rm_V> 15$ mag, we find a clear scaling relation between LECRIR and the local star formation rate (SFR): $log_{10}\zeta = (1.4\pm 0.70)log_{10}\mathrm{SFR} + (-10.5\pm 2.9)$ (Fig. \ref{fig:lecrir_map}c). This correlation provides strong evidence that LECRs are produced in situ by star-forming activities rather than originating from the broader Galactic cosmic-ray population.

    \item $\gamma$-ray Validation: Analysis of \textit{Fermi}-LAT data reveals that the 0.1--1 GeV cosmic-ray proton flux toward Regions 3 and 4 is $\sim$1.5--2.5 times higher than in Regions 1 and 2 (Fig. \ref{fig:lecrir_map}c and Fig. \ref{fig:fermiemi}b). This spatial enhancement in $\gamma$-ray emission independently corroborates our HINSA-derived LECRIR measurements.
    
\end{enumerate}

By synthesizing radio spectroscopy -- acting as an astronomical analogue to in situ particle measurements -- with $\gamma$-ray observations and \textit{Voyager} data, we provide compelling evidence for the local origin of low-energy cosmic rays. Future observations from the Square Kilometre Array (SKA) and next-generation $\gamma$-ray observatories will further refine our understanding of the origin and transport of cosmic rays across the interstellar medium.

\begin{acknowledgments}
This work was supported by the National Natural Science Foundation of China (grant Nos. 12588202, 12473023, 125B2059). N.-Y. T. acknowledges support by the University Annual Scientific Research Plan of Anhui Province (Nos. 2023AH030052). D. L. acknowledges support from the New Cornerstone Science Foundation. This work is supported by the Leading Innovation and Entrepreneurship Team of Zhejiang Province of China (Grant No. 2023R01008). This work was carried out in part at the Jet Propulsion Laboratory, which is operated by the California Institute of Technology under a contract with the National Aeronautics and Space Administration (80NM0018D0004). Rui-zhi Yang is supported by the NSFC under grant 12393854, 12041305 and 12588101,  Rui-zhi Yang gratefully acknowledge the support of Cyrus Chun Ying Tang Foundations and of the studio of Academician Zhao Zhengguo, Deep Space Exploration Laboratory. Bing Liu is funded by the Basic Research Program of Jiangsu (BK20252108). Tao-Chung Ching is a Jansky Fellow of the National Radio Astronomy Observatory.  

We appreciate the referee’s constructive suggestions, which have significantly improved the quality of this paper. We thank M. H. Heyer for providing CO(1-0) data toward Orion molecular cloud. This work has used the data from the Five-hundred-meter Aperture Spherical radio Telescope (FAST).  FAST is a Chinese national mega-science facility, operated by the National Astronomical Observatories of Chinese Academy of Sciences (NAOC).
\end{acknowledgments}

\appendix

\section{Data Analysis of High Energy Gamma-rays}
\label{sec:gamma-ray-analysis}
{\sl Although the gamma-rays in the MeV range are more closely related to LECRs, the poor angular resolution of Fermi-LAT in this energy range fails to provide sufficient information.} 
We thus used Fermi-LAT gamma-ray events from 0.1 to 500 GeV to study the spatial distribution of gamma-rays. The gamma-ray counts map is shown in Fig. \ref{fig:fermiemi}(a).

\subsection{Spatial Templates for Fermi-LAT Data Analysis}
\label{sec:template}
Assuming that gamma rays trace the distribution of the molecular gas, we generated spatial templates of the Orion molecular cloud for use in the analysis of \textit{Fermi}-LAT data. We utilized data from the Planck dust opacity map \citep{2014Planck} to trace the distribution of total gas column density. We adopted Eq. (4) from \citep{2014Planck}, in which the dust opacity $\tau_{\rm M}$ is represented as a function of wavelength $\lambda$, $\tau_{\rm M}(\lambda) = \left(\frac{\tau_{\rm D}(\lambda)}{N_{\rm H}}\right)^{\rm dust}[N_{{\rm HI}}+2X_{\rm CO}W_{\rm CO}].$ The term $(\tau_{\rm D}/N_{\rm H})^{\rm dust}$  represents the reference dust emissivity determined in low-$N_{\rm H}$ regions, while $X_{\rm CO}=N_{\rm H_2}/W_{\rm CO}$ is the conversion factor between the integrated brightness temperature of the CO ($W_{\rm CO}$) and the density of molecular hydrogen ($N_{\rm H_2}$). The total gas density ($N_{\rm H}$) can thus be estimated as $N_{\rm H} =N_{\rm {HI}}+2 N_{\rm H_2} = \tau_{\rm M}(\lambda)\left[\left(\frac{\tau_{\rm D}(\lambda)}{N_{\rm H}}\right)^{\rm dust}\right]^{-1}.$ The gas column density maps  were derived by using the 353~GHz dust emissivity value \citep{2014Planck} $(\tau_{\rm D}/N_{\rm H})^{\rm dust}_{353{\rm~GHz}}=(8.4\pm3.0)\times10^{-27}$~cm$^2$. We then used the derived gas column density map to generate spatial distribution templates of gamma-ray emission from the Orion molecular cloud. To find whether the gamma-ray emission from different parts of Orion is different in nature, we divided the region into 5 regions. The division of the regions is consistent with that of Fig. \ref{fig:lecrir_map}(a). However, the poor angular resolution of the gamma-ray data can not support the separation of the five regions, especially in the MeV range. So we combined Region 1 and 2 (hereafter Region 1 \& 2), as well as Region 3 and 4 (hereafter Region 3 \& 4), for joint fitting. To reduce the influence of the gas from the background, the pixels with a gas column density lower than $8\times10^{21}~\rm cm^{-2}$ were set to zero.

\subsection{Spatial Analysis of Fermi-LAT Data}
 Standard binned analysis was performed for each of the two ROIs. We followed the official tutorial of binned likelihood analysis\footnote{\url{https://fermi.gsfc.nasa.gov/ssc/data/analysis/scitools/binned_likelihood_tutorial.html}}. The source models were generated using make4FGLxml.py\footnote{\url{https://fermi.gsfc.nasa.gov/ssc/data/analysis/user/make4FGLxml.py}}, which included sources from the LAT {\bf 14}-year Source Catalog (4FGL-DR4, \cite{4FGL}) within the ROI enlarged by 10$^{\circ}$, and the isotropic emission background (iso\_P8R3\_SOURCE\_V3\_v1.txt). The spectral parameters of sources within 10\deg\ from the center and the normalization factor of the diffuse and isotropic background were set free. We then added Region 1 \& 2 and Region 3 \& 4 spatial templates in the source model of the analysis centered on Orion molecular cloud to model their gamma-ray emission. Meanwhile, the gas column density maps within each ROI enlarged by 10$^{\circ}$ in which the Region 1 \& 2 and Region 3 \& 4, are set as the background templates instead of the Galactic interstellar emission model (\texttt{gll\_iem\_v07.fits}) provided by the Fermi-LAT Collaboration. Besides the gamma rays produced from the pion-decay process induced by the inelastic collision between CR protons and ambient gas, the diffuse emission also includes the inverse Compton (IC) scattering of CR electrons in the interstellar radiation fields (ISRFs).  Thus, we generated the diffuse IC emission template using GALPROP\footnote{\url{http://galprop.stanford.edu/webrun/}} \citep{galprop} with the parameter set $^SY ^Z4^R30^T150^C2$ in \citet{fermi_diffuse_old}.

\subsection{Spectral Analysis of Fermi-LAT Data}
Based on the two analyses in the last section, we continued to perform spectral analyses. To derive the gamma-ray spectrum of the three regions of Orion, we divided the data from $100\rm~MeV$ to $500\rm ~GeV$ into 9 energy bins evenly distributed in logarithmic space. For the data in each energy bin, we conducted a likelihood analysis and derived the gamma-ray flux in each energy band. The derived gamma-ray spectra normalized to gamma-ray emissivity per hydrogen atom are shown in Fig. \ref{fig:fermiemi}. For the last two energy bins, likelihood analysis could not be successfully performed due to insufficient data size. Consequently, only the results from the first seven bins were utilized.

\subsection{Derivation of High Energy CR Spectra}
\begin{figure*}
\begin{center}  
 \includegraphics[width=0.5\textwidth]{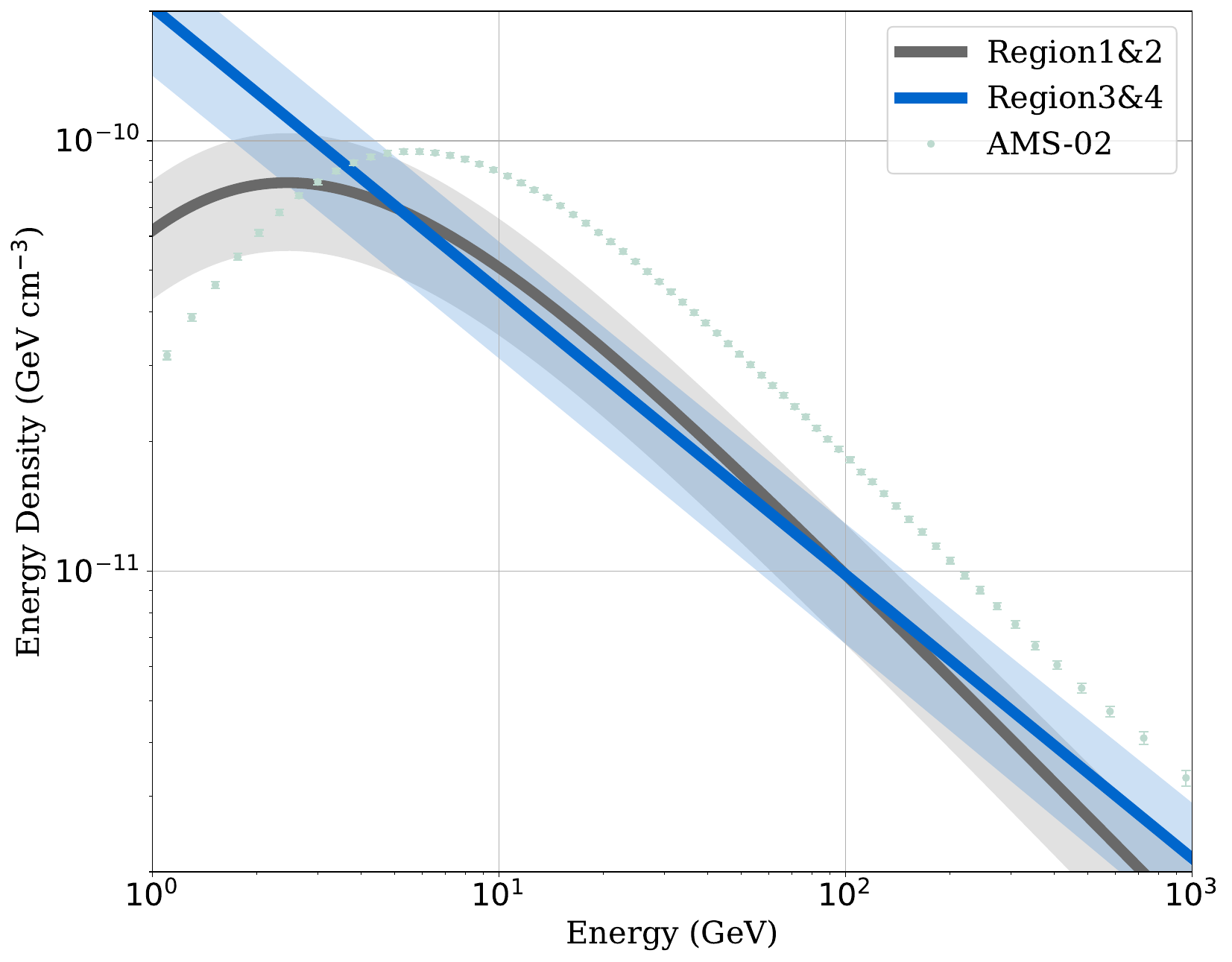}
\caption{Cosmic ray energy densities in different regions of the Orion molecular cloud. The cosmic ray energy densities in Region 1 \& 2 and Region 3 \& 4 are shown with grey and blue colors, respectively. The data points indicate the local CR proton energy density measured by AMS-02 \citep{Aguilar15}.}
\label{fig:fermicr} 
\end{center}
\end{figure*}

The observed pion-bump feature suggests that gamma-ray emission in Orion primarily originates from proton-proton interactions between CR protons and interstellar gas. We then assumed kinetic energy of the CR protons follow the distribution function: $F=f_i\left(E+E_b\right)^{\Gamma}$, where $F$ is the flux density of protons in the unit of $\mathrm{cm}^{-2}\, \mathrm{s}^{-1}\,\mathrm{GeV}^{-1}$,  $E$ is the proton kinetic energy in the unit of $~\rm GeV$, $E_b$ is the break energy in the unit of $~\rm GeV$, and $\Gamma$ stands for the spectral index. Under this assumption, we applied the gamma-ray production cross-section from \cite{2014pp} to conduct a likelihood analysis of the gamma-ray data. We further determined the gas mass in the interstellar template (Section \ref{sec:template}) to derive absolute CR fluxes. We found that to best fit the data, the $E_b$ of region 1 \& 2 and region 3 \& 4 are approximately $1.0~\rm GeV$ and $0~\rm GeV$, respectively. The derived best-fit CR proton spectral index $\Gamma$ is 2.82$\pm$0.03 for region 1 \& 2, and 2.66$\pm$0.03 for region 3 \& 4. The derived CR spectra are shown in Extended Data Figure \ref{fig:fermicr}. The gamma-ray emissivities per hydrogen atom derived from the fitted cosmic-ray spectra are represented by dashed lines in Fig. \ref{fig:fermiemi}(b).

\bibliography{sample701}{}
\bibliographystyle{aasjournalv7}



\end{document}